\documentclass[aps,prd,twocolumn,reprint,superscriptaddress,nofootinbib,floatfix]{revtex4-2}
\pdfoutput=1

\usepackage{amsmath,amssymb,amsfonts}
\usepackage{bbm}
\usepackage{graphicx}
\usepackage[utf8]{inputenc}
\usepackage[T2A]{fontenc}
\usepackage[colorlinks]{hyperref}
\usepackage[english]{babel}
\usepackage[usenames]{color}
\usepackage{slashed}

\definecolor{darkorange}{rgb}{1,0.549,0}

\usepackage[normalem]{ulem}

\newcommand{\RG}{{\small RG}}
\newcommand{\GR}{{\small GR}}
\newcommand{\CDT}{{\small CDT}}
\newcommand{\QED}{{\small QED}}
\newcommand{\YM}{{\small YM}}
\newcommand{\eg}{{\textit{e.g.}}}

\newcommand{\mans}{\ensuremath{\mathfrak{s}}}
\newcommand{\mant}{\ensuremath{\mathfrak{t}}}

\newcommand{\con}[2]{\ensuremath{\Gamma^{#1}_{\phantom{#1}#2}}}
\newcommand{\anticon}[2]{\ensuremath{\reflectbox{$\Gamma$}_{#1}^{\phantom{#1}#2}}}
\newcommand{\riem}[2]{\ensuremath{R_{#1}^{\phantom{#1}#2}}}
\newcommand{\deltatensor}[2]{\ensuremath{\delta_{#1}^{\phantom{#1}#2}}}
\newcommand{\ric}{R}
\newcommand{\invric}{\reflectbox{\ric}}
\newcommand{\De}{\text{Д}}
\newcommand{\CD}{D}
\newcommand{\bCD}{\bar{D}}
\newcommand{\dGamma}[2]{\ensuremath{{\delta\Gamma}^{#1}_{\phantom{#1}#2}}}
\newcommand{\WdW}[4]{\ensuremath{{\mathcal{G}}^{\phantom{#1}#2\phantom{#3}#4}_{#1\phantom{#2}#3\phantom{#4}}}}
\newcommand{\alphag}{\ensuremath{\alpha_g}}
\newcommand{\CC}{\ensuremath{\Lambda}}
\newcommand{\metric}{g}
\newcommand{\Ctensor}[2]{\ensuremath{C_{#1}^{\phantom{#1}#2}}}
\newcommand{\prop}[4]{\ensuremath{\mathfrak G_{\phantom{#1}#2\phantom{#3}#4}^{#1\phantom{#2}#3}}}
\newcommand{\propTS}[5]{\ensuremath{\mathcal T_{#5\, \phantom{#1}#2\phantom{#3}#4}^{\phantom{#5\,}#1\phantom{#2}#3}}}

\allowdisplaybreaks

\begin{document}

\title{Scattering amplitudes in affine gravity}
\author{Benjamin Knorr\,\href{https://orcid.org/0000-0001-6700-6501}{\protect \includegraphics[scale=.07]{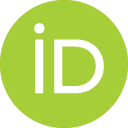}}\,}
\email[]{bknorr@perimeterinstitute.ca}
\affiliation{
Perimeter Institute for Theoretical Physics, 31 Caroline St. N., Waterloo, ON N2L 2Y5, Canada
}
\author{Chris Ripken\,\href{https://orcid.org/0000-0003-2545-5047}{\protect \includegraphics[scale=.07]{ORCIDiD_icon128x128.png}}\,}
\email[]{aripken@uni-mainz.de}
\affiliation{
Institute of Physics (THEP), University of Mainz, Staudingerweg 7, 55128 Mainz, Germany
}

\begin{abstract}
Affine gravity is a connection-based formulation of gravity that does not involve a metric. After a review of basic properties of affine gravity, we compute the tree-level scattering amplitude of scalar particles interacting gravitationally via the connection in a curved spacetime. We find that, while classically equivalent to general relativity, affine gravity differs from metric quantum gravity.
\end{abstract}

\maketitle

\section{Introduction}\label{sec:intro}
Even more than 100 years after the formulation of General Relativity (\GR{}), and despite an enormous effort, we still lack a consistent and experimentally verified theory of quantum gravity. The standard approach that works excellently for the Standard Model of particle physics, perturbation theory, fails to provide an acceptable quantisation of \GR{} beyond the framework of effective field theory. The reason is that every new loop order introduces new interaction terms which are not part of the original action \cite{'tHooft:1974bx}, and with it, new coupling constants appear that have to be fixed by some experiment. In pure gravity, this happens for the first time at the two-loop level with the famous Goroff-Sagnotti term \cite{Goroff:1985sz, Goroff:1985th, vandeVen:1991gw}, whereas with matter, the first occurrence is at one-loop order. One thus finds oneself in the situation that in principle one would have to conduct infinitely many experiments to uniquely specify the theory, and thus any predictivity is lost. This behaviour is also expected from naive power counting: the coupling of \GR{}, Newton's constant $G_N$, has a mass dimension of $-2$ in four spacetime dimensions.

Many different ideas have been put forward to circumvent this problem in one way or another. The most conservative possibility to solve this problem is that gravity can be renormalised in a non-perturbative way. Approaches in this direction include the Asymptotic Safety program \cite{Reuter:1996cp, Reuter:2012id, Reuter:2019byg, Reichert:2020mja, Bonanno:2020bil}, Causal Dynamical Triangulations (\CDT{}) \cite{Ambjorn:2012jv, Loll:2019rdj}, and loop quantum gravity \cite{Bahr:2011yc, Rovelli:2014ssa, Steinhaus:2020lgb}. Closely related are more geometry-based approaches like causal sets \cite{Surya:2019ndm}, with the disadvantage that the connection to standard quantum field theory is less clear. An alternative is to give up on some of the symmetries, and for example break Lorentz symmetry at small scales \cite{Horava:2009uw, Barvinsky:2017kob, Steinwachs:2020jkj, Knorr:2018fdu, Eichhorn:2019ybe}.

A central question in all of these approaches is what the fundamental degrees of freedom are. For example, whereas in general Asymptotic Safety is agnostic about this, an overwhelming majority of research conducted in this area is based on the metric as the fundamental field, with the occasional exception based, \eg{}, on the tetrad formalism \cite{Dona:2012am}.
\CDT{} is based on purely geometric objects, on which the diffeomorphism group acts trivially. In loop quantum gravity, a central role is played by holonomies. Usually all these different choices have in common that at the classical level they are equivalent to \GR{}.

A particular choice of degrees of freedom has not received a lot of attention, even though it was first discussed quite early -- the purely affine formulation, based only on the use of an affine connection. Originally discussed by Eddington and Schr\"odinger \cite{10.2307/20488479, 10.2307/20488482, schroedinger_1985}, most investigations involving the connection as a fundamental degree of freedom are carried out in a formulations where the relevant connection lives in a related gauge bundle, \eg\ $SO(3)$ \cite{Krasnov:2011pp, Krasnov:2012pd, Delfino:2012zy, Delfino:2012aj, Groh:2013oaa, Krasnov:2017ygw}, or $SO(1,3)$ \cite{Ashtekar:2015scholarpedia}. On paper, the purely affine theory has many interesting properties. First, it is much closer in spirit to the other fundamental forces in nature. In this picture, there is a unification of the description: forces are mediated by a connection (be it gravitational, strong or electro-weak), whereas matter is described by fermions and the Higgs boson. Second, the classical theory is completely equivalent to \GR{} with a cosmological constant, however the latter appears as an integration constant similar to the case of unimodular gravity. A key difference to the metric formulation is that the equation of motion for the connection is \emph{cubic} instead of non-polynomial, once again emphasising the similarity to classical Yang-Mills (\YM{}) theory. Third, while having a metric is certainly useful, it is not necessary to describe, \eg{}, geodesic motion, which solely depends on the connection. Fourth, the implementation of torsion degrees of freedom is completely straightforward. While torsion is usually not considered (partially because in \GR{}, torsion is not dynamical), it is a question of experiment whether our universe has torsion degrees of freedom or not. Finally, and maybe most promising from the viewpoint of quantisation, the coupling constant in the affine theory is dimensionless. This opens up the possibility of a perturbative quantisation of gravity without making other sacrifices like assumptions on extra fields, dimensions or breaking of symmetries.

In this paper, we take all these reasons as a motivation to study affine gravity from a modern perspective. With the anticipation that not many readers will be familiar with the theory, we provide an overview of the basics in \autoref{sec:basics}, systematising some of the earlier literature. In particular, we will discuss its action and equation of motion, the symmetry structure of the theory, as well as the coupling to matter. With this in place, in \autoref{sec:amplitude} we derive the scattering amplitude of a two-scalar-to-two-scalar process mediated by the gravitational connection and discuss the differences to \GR{}. This will be done on a curved spacetime which fulfils the equations of motion. We then close with a summary of the results and an outlook in \autoref{sec:summary}.

\section{Basics of affine gravity}\label{sec:basics}

In this section we present the foundations of affine gravity. Since we do not have access to a metric, we will be very careful with the index positions of all tensors in any expression. To make matters simpler, in this work we restrict ourselves to a symmetric connection so that the torsion vanishes:
\begin{equation}
 \con{\alpha}{\mu\nu} = \con{\alpha}{(\mu\nu)} \, .
\end{equation}
The brackets indicate the symmetrisation of the enclosed indices normalised to unit strength. The Riemann and symmetrised Ricci tensors in terms of this connection are then
\begin{equation}\label{eq:riemofcon}
\begin{aligned}
 \riem{\mu\nu\rho}{\alpha} &= \con{\alpha}{\nu\beta} \con{\beta}{\mu\rho} - \con{\alpha}{\mu\beta} \con{\beta}{\nu\rho} - \partial_\mu \con{\alpha}{\nu\rho} + \partial_\nu \con{\alpha}{\mu\rho} \, , \\
 \ric_{\mu\nu} &=  \riem{(\mu|\alpha|\nu)}{\alpha} \, .
\end{aligned}
\end{equation}
Vertical bars indicate that the enclosed indices are not symmetrised over.
In affine gravity we will generally have to assume that the symmetrised Ricci tensor is non-degenerate so that it has a multiplicative inverse which we denote by \invric{},
\begin{equation}
 \invric^{\mu\alpha} \ric_{\alpha\nu} = \deltatensor{\nu}{\mu} \, .
\end{equation}
Effectively, $\ric{}$ and $\invric{}$ will be used to lower and raise indices, with the difference to the metric case that in general their covariant derivatives are not zero. In this sense, it is akin to not using the Levi-Civita connection in the metric case. From the invertibility requirement of the symmetrised Ricci tensor we conclude that the limit of having a flat spacetime will have to be implemented very carefully. Finally, we can define the equivalent of a d'Alembertian operator, which we will call \De{}, by
\begin{equation}
 \De = -\CD_\mu \invric^{\mu\nu} \CD_\nu \, .
\end{equation}
In this, $\CD{}$ is the covariant derivative associated to the connection \con{}{}. Notice that, in contrast to the d'Alembertian, \De{} has a vanishing mass dimension, so its eigenvalues are pure numbers. We will see that this is a general property of the affine setup: the mass dimension of an expression is related to the number of its free indices.

\subsection{Action and equation of motion}

Let us first discuss the classical action of affine gravity. In a general spacetime dimension $d$, it is given by
\begin{equation}\label{eq:SEddington}
 S^\text{Edd} = \frac{1}{4\alphag} \int \text{d}^dx \, \sqrt{-\det \ric_{\mu\nu}} \, .
\end{equation}
Several remarks are in order. First, the coupling $\alphag$ is dimensionless for all $d$, and the nomenclature is chosen to emphasise the closeness to \YM{} theory. Second, it is non-polynomial in the Ricci curvature tensor, so eventually we have to carefully assess whether the theory possesses any ghost or tachyonic modes. Finally, for the action to be real, we have to require that the Ricci tensor retains its signature. This is similar to the requirement in metric gravity that the metric has a fixed signature, and will constrain the domain of integration of the path integral.

To derive the equations of motion, we perform a standard variation of the action with respect to the connection \con{}{}. After some algebra, and using that $\ric{}$ is non-degenerate, one can show that the equation of motion has the very compact form
\begin{equation}\label{eq:eom}
 \CD_\mu \ric_{\alpha\beta} = 0 \, .
\end{equation}
Recalling the dependence of the Ricci tensor on the connection, \eqref{eq:riemofcon}, this shows that the equation of motion is a cubic second-order partial differential equation, similar to that of \YM{} theory. One can use the equation of motion to show that the connection is the Levi-Civita connection with respect to the Ricci tensor,
\begin{equation}
 \con{\mu}{\alpha\beta} = \frac{1}{2} \invric^{\mu\nu} \left( \partial_\alpha \ric_{\nu\beta} + \partial_\beta \ric_{\nu\alpha} - \partial_\nu \ric_{\alpha\beta} \right) \, .
\end{equation}

The equation \eqref{eq:eom} is equivalent to \GR{} with a cosmological constant. It states that the Ricci tensor is covariantly constant, so we can write suggestively
\begin{equation}\label{eq:EEsol}
 \ric_{\mu\nu} = \frac{2\CC}{d-2} \, \metric_{\mu\nu} \, ,
\end{equation}
where \CC{} is a constant of mass dimension two, $\metric{}$ is a covariantly constant tensor of mass dimension zero, and the normalisation is chosen to allow a simple relation to \GR{}. Clearly, \eqref{eq:EEsol} is just Einstein's equation with cosmological constant \CC{} and metric $\metric{}$. With this, we can relate the coupling \alphag{} to Newton's constant $G_N$ and the cosmological constant. Requiring that the classical Einstein-Hilbert action of \GR{},
\begin{equation}
 S^\text{GR} = \frac{1}{16\pi G_N} \int \text{d}^dx \sqrt{-\det \metric_{\mu\nu}} \, \left( \ric{} - 2\CC{} \right) \, ,
\end{equation}
agrees with the Eddington action \eqref{eq:SEddington} on the solution of the equation of motion \eqref{eq:EEsol}, we find
\begin{equation}
 \alphag = 2\pi G_N \left( \frac{2\CC{}}{d-2} \right)^{\frac{d-2}{2}} \stackrel{d=4}{=} 2\pi G_N \CC{} \, .
\end{equation}
The observed value of \alphag{} is extremely tiny (of the order of $10^{-120}$), so that we can expect excellent convergence of any perturbative treatment in this coupling.

Let us stress the conceptual difference of the origin of Newton's constant and the cosmological constant in \GR{} and affine gravity. In \GR{}, both constants are part of the action, and thus are fundamental parameters of the theory. By contrast, in affine gravity, only the coupling \alphag{} is part of the definition of the action. The cosmological constant arises as a constant of integration in the solution of the equation of motion, similar to unimodular gravity. In this way, the tiny observed value of the cosmological constant is just a statement about what initial condition is realised in the universe. Trying to explain its small value from first principles is akin to asking for an explanation of why a harmonic oscillator starts at the exact position where it starts.

\subsection{Coupling to matter}

We will now discuss some aspects of the coupling to matter in affine gravity. Since we do not have access to a metric off-shell, there are a few differences with regards to the metric case. The first observation is that the measure is already dimensionless, in contradistinction to the metric case. This means that
\begin{equation}
 \int \text{d}^d x \, \sqrt{-\det \ric_{\mu\nu}} \, ,
\end{equation}
is a pure number and carries the interpretation of a volume in units of the curvature. As a consequence, any Lagrangian density that is included will have to be dimensionless. This changes the mass dimension of some fields compared to the standard case. For example, writing down a kinetic term for a scalar field $\phi$,
\begin{equation}
 S^\text{scal} = \frac{1}{2} \int \text{d}^d x \, \sqrt{-\det \ric_{\mu\nu}} \, \invric^{\alpha\beta} (\CD_\alpha \phi) (\CD_\beta \phi) \, ,
\end{equation}
it is evident that the scalar field has zero mass dimensions. By contrast, a gauge field continues to have a mass dimension of one, and a kinetic term for the photon reads
\begin{equation} \label{eq:Sphot}
 S^\text{phot} = \frac{1}{4} \int \text{d}^d x \, \sqrt{-\det \ric_{\mu\nu}} \, \invric^{\alpha\gamma} \invric^{\beta\delta} F_{\alpha\beta} F_{\gamma\delta} \, ,
\end{equation}
where $F$ is the field strength tensor.

For fermions, the situation is a bit more involved. A first step in the construction of the spinor bundle would be to define the Clifford algebra in terms of the Ricci tensor,
\begin{equation}
 \left\{ \gamma_\mu, \gamma_\nu \right\} = 2\ric_{\mu\nu} \, \mathbbm 1 \, .
\end{equation}
Here, $\gamma_\mu$ are the mass dimension one gamma matrices, and the matrix $\mathbbm 1$ is the identity in Dirac space.
However, the full construction of the spinor bundle in the absence of a Levi-Civita connection goes beyond the scope of this work.

We will now discuss how to obtain the mass dimension of any object. By definition, the partial derivative has mass dimension one, and comes with one lower spacetime index. To be consistent, any (gauge) connection must have the same mass dimension so that covariant derivatives have a uniform mass dimension. From this it follows that any curvature tensor has mass dimension two, and the inverse Ricci tensor then has a mass dimension of minus two. With these facts, it is straightforward to convince oneself that the mass dimension of any tensor is given by the difference of the number of lower spacetime indices minus the number of upper spacetime indices. In particular, any spacetime scalar has vanishing mass dimension. The interpretation of this compared to standard quantum field theory with a metric is that in affine gravity the fields are measured in units of the spacetime curvature (which is on-shell equivalent to measuring in units of the cosmological constant).

This has two very unconventional consequences: all conceivable coupling constants must have mass dimension zero, so naive power counting stops to be applicable. This might be a potential roadblock to the perturbative quantisation of the theory, but more investigations are necessary. In particular, on solutions to the equation of motion, standard quantum field theory, including its power counting, must emerge under a suitable rescaling of the fields. Second, no ad-hoc scale is introduced at any level. Every measurement of a dimensionful quantity is relational in the sense that it will be expressed in units of some other dimensionful quantity of the same physical system.

\subsection{Symmetry structure}\label{sec:symstruc}
We proceed our discussion of affine gravity by investigating its symmetry properties. Naturally, any action for affine gravity that we propose should be diffeomorphism-invariant. Infinitesimally, this is made explicit by the condition
\begin{equation}
			S^{\text{Edd}}(\Gamma+ \mathcal{L}_\epsilon \Gamma)
	=
			S^{\text{Edd}}(\Gamma)
	\,\text{,}
\end{equation}
where $\mathcal{L}_\epsilon$ denotes the Lie derivative with respect to the vector field $\epsilon$ that generates a diffeomorphism.
The Lie derivative acting on the connection $\Gamma$ is given by
\begin{equation}
			{(\mathcal{L}_\epsilon \Gamma)^\alpha}_{\mu\nu}
	=
			\CD_\mu \CD_\nu	\epsilon^\alpha	+	\riem{\mu\beta\nu}{\alpha}	\epsilon^\beta
	\,\text{.}
\end{equation}
In order to compute the connection propagator in affine gravity, it is necessary to gauge-fix the diffeomorphism symmetry. For this, we employ the background field method. Here, the connection is split into a background connection $\bar\Gamma$ and a fluctuation $\dGamma{}{}$,
\begin{equation}
			\Gamma
	=
			\bar \Gamma + \dGamma{}{}
	\,\text{.}
\end{equation}
The first step in the gauge-fixing procedure is to define a DeWitt inner product on the space of fluctuations \cite{DeWitt:1967yk}. For this, we employ the symmetrised Ricci tensor with respect to the background connection. The most general ultra-local inner product reads
\begin{equation}
			\left\langle
				\dGamma{}{}
			,
				\dGamma{}{}
			\right\rangle
	=
			\int	\mathrm{d}^d x	\sqrt{-\det \bar{R}_{\kappa\lambda}}	\, \dGamma{\alpha}{\mu\nu}	\WdW{\alpha}{\mu\nu}{\beta}{\rho\sigma}	\dGamma{\beta}{\rho\sigma}
	\,\text{,}
\end{equation}
where the tensor $\WdW{}{}{}{}$ is given by
\begin{equation}
\begin{aligned}
 \WdW{\alpha}{\mu\nu}{\beta}{\rho\sigma} &= \bar\ric_{\alpha\beta} \bar\invric^{\mu(\rho} \bar\invric^{\sigma)\nu} \\
 &\qquad + \frac{\beta_1}{2} \left( \deltatensor{\alpha}{\mu} \deltatensor{\beta}{(\rho} \bar\invric^{\sigma)\nu} + \deltatensor{\alpha}{\nu} \deltatensor{\beta}{(\rho} \bar\invric^{\sigma)\mu} \right) \\
 &\qquad + \beta_2 \, \bar\ric_{\alpha\beta} \bar\invric^{\mu\nu} \bar\invric^{\rho\sigma} \\
 &\qquad + \frac{\beta_3}{2} \left( \bar\invric^{\mu\nu} \deltatensor{\alpha}{(\rho} \deltatensor{\beta}{\sigma)} + \bar\invric^{\rho\sigma} \deltatensor{\alpha}{(\mu} \deltatensor{\beta}{\nu)} \right) \\
 &\qquad + \beta_4 \, \deltatensor{\alpha}{(\rho} \bar\invric^{\sigma)(\mu} \deltatensor{\beta}{\nu)} \, .
\end{aligned}
\end{equation}
The parameters $\{\beta_i\}$ are gauge parameters. The DeWitt metric is invertible for most choices of $\{\beta_i\}$, except for a few isolated cases.

The gauge fixing action is now constructed via,
\begin{equation}
			S^{\text{gf}}
	=
			\frac{1}{2\alpha}	\int \text{d}^dx	\sqrt{-\det \bar\ric_{\mu\nu}}	\,	\bar\invric^{\gamma\delta}	F_\gamma[\dGamma{}{},\bar\Gamma]	F_\delta[\dGamma{}{},\bar\Gamma]
	\,\text{,}
\end{equation}
where the gauge fixing condition is implemented by the functional $F$ using the deWitt metric,
\begin{equation}
			F_\gamma[\dGamma{}{},\bar\Gamma]
	=
			\WdW{\alpha}{\mu\nu}{\beta}{\rho\sigma}\left(
				\deltatensor{\gamma}{\beta}	\bCD_\sigma	\bCD_\rho	
			+	{\bar R_{\rho\gamma\sigma}}^{\phantom{\rho\gamma\sigma}\beta}	
			\right)	\dGamma{\alpha}{\mu\nu}
			\,.
\end{equation}

\subsection{Connection to metric theory}

In this section we explain how to recover standard metric theory from affine gravity in cases where we can define a metric, \eg{}, via the equation of motion \eqref{eq:EEsol}. As we have observed earlier, mass dimensions of fields and couplings in the theory are different from the standard case, so all fields and couplings have to be rescaled by appropriate powers of the only scale available, the cosmological constant, introduced by the equation of motion. For example, a scalar field in the affine theory is dimensionless,
\begin{equation}
 [\phi] = 0 \, ,
\end{equation}
whereas in a standard metric theory, it has a mass dimension
\begin{equation}
 [\phi_\metric] = \frac{d-2}{2} \, .
\end{equation}
Here the subscript $g$ indicates that the quantity belongs to a metric theory. Accordingly, up to constant rescalings by numbers, the two fields are related by
\begin{equation}\label{eq:scalarfieldrescaling}
 \phi = \CC^{\frac{2-d}{2}} \phi_\metric \, .
\end{equation}
From this, we can derive the scaling of any coupling constant. For example, consider an term in the action of the structural form
\begin{equation}
 \kappa \int \text{d}^dx \sqrt{-\det \ric_{\mu\nu}} \, \De{}^m \, \phi^n \, ,
\end{equation}
where it is understood that the derivatives are distributed in some way among the fields, and $\kappa$ is a coupling constant. Replacing the curvatures via \eqref{eq:EEsol} and the scalar fields by \eqref{eq:scalarfieldrescaling} while dropping overall constants, this term becomes
\begin{equation}
 \kappa \, \CC^{d-m+\frac{2-d}{2}n} \int \text{d}^dx \sqrt{-\det \metric_{\mu\nu}} \, \Delta^m \, \phi_\metric^n \, .
\end{equation}
Once again, the standard Laplacian $\Delta=-g^{\mu\nu}D_\mu D_\nu$ is understood to be distributed among the scalar fields in the same way as above. From this, we can read off the necessary rescaling of the dimensionless coupling $\kappa$ to define its dimensionful metric counterpart $\kappa_\metric$,
\begin{equation}
 \kappa = \CC^{-d+m-\frac{2-d}{2}n} \kappa_\metric \, .
\end{equation}
The same algorithm applies to relate other matter fields to their metric counterparts. This emphasises again the interpretation that all quantities in affine gravity are measured in units relative to the only scale spacetime itself gives us in the theory, the cosmological constant.

\subsection{Synopsis of previous results}
In this section, we collect results on affine gravity that have been previously obtained. We will first consider classical aspects of the theory. Next, we will briefly review known quantum aspects.

We start our discussion by noting that the affine formulation, the metric formulation, and metric-affine formulations such as the Palatini formalism can all be shown to be classically equivalent \cite{Ferraris:1981hr,Ferraris:1982ez}. The proof of this equivalence shows that the metric degree of freedom arises in the affine formulation as the canonical conjugate of the symmetric part of the Ricci tensor. The canonical structure of the affine theory was discussed in \cite{Kijowski:2004bj}.

The identification of the metric as the symmetric part of the Ricci tensor can be generalised to unify gravity and $U(1)$ gauge fields \cite{Ferraris:1981hr,Ferraris:1982fa,Poplawski:2006zr,Poplawski:2007cc}. Here, the electromagnetic field strength tensor $F_{\mu\nu}$ is identified with the antisymmetric part of the Ricci tensor. Variation of the action \eqref{eq:Sphot} with respect to the connection then gives both the Einstein equations with cosmological constant, and the Maxwell equations.

An initial study of the quantum properties of affine gravity was performed in \cite{Martellini:1984ec}. In this work, the one-loop renormalisation group (\RG{}) running of the coupling $\alphag$ was computed. This $\beta$-function exhibits an infrared-attractive fixed point at $\alphag=0$, similar to the electromagnetic fine-structure constant in \QED{}. Since $\alphag$ is related to the cosmological constant, such a fixed point is particularly favoured by the small observed value of $\CC$. We note that \cite{Martellini:1984ec} first recasts the Eddington action \eqref{eq:SEddington} into an on-shell equivalent, polynomial action. While the quantum properties of this theory are certainly appealing, it is presently unknown whether the $\beta$-function of the original theory \eqref{eq:SEddington} agrees with these findings.

\section{Scattering amplitude}\label{sec:amplitude}

While affine gravity agrees with \GR{} when it comes to the equation of motion, the off-shell degrees of freedom are expected to differ, so that their respective quantised versions potentially give rise to different results. A straightforward way to probe this is the calculation of a tree-level scattering amplitude. Since pure gravitational scattering is technically challenging, we will discuss a gravity-mediated scalar scattering of a single minimally coupled massless scalar field.

\subsection{Setup}

Our ansatz for the action is
\begin{equation}
 S = \int \text{d}^dx \, \sqrt{-\det \ric_{\mu\nu}} \left[ \frac{1}{4\alphag} + \frac{1}{2} \invric^{\alpha\beta} (\CD_\alpha\phi)(\CD_\beta\phi) \right] \, .
\end{equation}
For the calculation of the amplitude of the scattering process $\phi\phi\to\phi\phi$, we need the connection propagator. Due to the underlying diffeomorphism symmetry, we have to employ a gauge fixing as discussed in \autoref{sec:symstruc}. To further simplify the calculation, we will use a covariantly constant background,
\begin{equation}
 \bCD_\mu \bar\ric_{\alpha\beta} = 0 \, ,
\end{equation}
and specify the Riemann tensor to be maximally symmetric,
\begin{equation}
 \bar{R}_{\mu\nu\rho}^{\phantom{\mu\nu\rho}\sigma} = \frac{1}{d-1} \left( \bar\ric_{\mu\rho} \deltatensor{\nu}{\sigma} - \bar\ric_{\nu\rho} \deltatensor{\mu}{\sigma} \right) \, .
\end{equation}
This is similar to, but clearly more general than, standard calculations of gravitationally mediated scattering amplitudes in a flat spacetime, and consistent since we do not discuss external gravitational legs. 
In order to ease the notation, we omit the bar to denote background quantities.

As a consequence of this choice of background, the on-shell condition for the scalar field $\phi$ simply reads
\begin{equation}
 \De \, \phi = 0 \, .
\end{equation}
Additionally, we assume that we can actually define asymptotic states on the given background. We expect that the technical progress of calculating scattering amplitudes in curved spacetimes that we put forward in the following will also be useful for metric-based calculations.

\subsection{Calculation}

We will now outline the calculation of the curved space scattering amplitude. Since we do not have access to momentum space, we will work in position space throughout the calculation.

The gravity-mediated amplitude consists of the contraction of two three-point vertices with a single connection propagator. In position space, a two-to-two amplitude is then an operator which acts on four different arguments. With this in mind, it is in practice easier to not take the derivatives with respect to the external fields, that is we formulate the amplitude in terms of its action on test fields. In that way, what we actually have to calculate is the contraction of the connection propagator $\mathfrak{G}$ sandwiched between two connection-energy-momentum tensors:
\begin{equation} \label{eq:amplitude_positionspace}
		\mathcal{A}(\{\phi_i\})
	=
		\int	\text{d}^dx \sqrt{-\det \ric_{\mu\nu}} \,  T_{\phi_1\phi_2} \,\,\mathfrak{G}\,\,	T_{\phi_3\phi_4}
	\,.
\end{equation}
We will call this object the amplitude functional, and the actual amplitude is the fourth functional derivative of it. The amplitude \eqref{eq:amplitude_positionspace} corresponds to the $s$-channel, and the labels $1,2$ indicate the incoming scalars whereas the labels $3,4$ indicate the outgoing scalars. The $t$- and $u$-channels follow by crossing symmetry.

With this general approach settled, the central object that we have to compute is the propagator, which is the inverse of the two-point function. Structurally, the propagator is a linear combination of functions of \De{} and tensor structures in such a way that the complete object is self-adjoint. We can choose a specific ordering of propagator functions and tensor structures to simplify the calculation. The choice that we implemented is to sort all propagator functions to the right of the respective tensor structure, so that the propagator reads
\begin{equation}\label{eq:propagator}
 \prop{\mu}{\alpha\beta}{\nu}{\kappa\lambda} = \sum_{\ell=1}^{32} \propTS{\mu}{\alpha\beta}{\nu}{\kappa\lambda}{\ell} G_\ell(\De{}) \, .
\end{equation}
Here, the $\propTS{}{}{}{}{\ell}$\! are 32 independent tensor structures that span a basis, and the $G_\ell$ are the propagator functions that have to be computed. 

To compute the propagator functions $\{G_\ell\}$, we demand that
\begin{equation}
 \frac{\delta^2 S^\text{Edd}}{\delta \con{\rho}{\tau\omega}\delta \con{\nu}{\kappa\lambda}} \prop{\nu}{\kappa\lambda}{\mu}{\alpha\beta} = \mathbf{i} \, \deltatensor{\rho}{\mu} \deltatensor{\alpha}{(\tau} \deltatensor{\beta}{\omega)} \, .
\end{equation}
To do this in practice, we act with this equation onto a test tensor \anticon{\mu}{\alpha\beta} with an index structure opposite to the connection, but the same symmetry. Next, all occurrences of \De{} are sorted to the right, so that they act first on \anticon{}{}. Finally, we symmetrise all remaining covariant derivatives to arrive at an expression in canonical form, where all tensor structures are independent. The propagator functions can then be read off.

With the propagator at hand, it remains to contract it with two first connection derivatives of the matter action to derive the scattering amplitude functional. Partial integration can be used to simplify the expression -- these are nothing else than the curved space equivalent of momentum conservation. Furthermore, the external test fields are assumed to be on-shell, so that any \De{} acting on an individual field vanishes. The full calculation is rather cumbersome, and we carried it out with the tensor algebra package \textit{xAct} \cite{xActwebpage, 2007CoPhC.177..640M, 2008CoPhC.179..597M, 2014CoPhC.185.1719N}. For the convenience of the reader, the complete Mathematica notebook detailing the calculation is provided as an ancillary file.

\subsection{Result}\label{sec:result}

To obtain a manifestly gauge-invariant result, the scattering amplitude has to be brought into a canonical form. Due to the structure of the action, we know that the amplitude functional can be written as
\begin{equation}\label{eq:amp_genform}
\begin{aligned}
 &\frac{\mathcal A}{4\alphag} = \int \text{d}^dx \sqrt{-\det\ric_{\alpha\beta}} \Big[\phi_1 \phi_2 a_0(\De) \phi_3 \phi_4 \\
 & + \invric^{\mu\nu} (\CD_\mu \phi_1) \phi_2 a_1(\De) (\CD_\nu \phi_3) \phi_4 \\
 & + \invric^{\mu\nu} \invric^{\rho\sigma} (\CD_{(\mu} \CD_{\rho)} \phi_1) \phi_2 a_2(\De) (\CD_{(\nu} \CD_{\sigma)}  \phi_3) \phi_4 \\
 & + \invric^{\mu\nu} \invric^{\rho\sigma} \invric^{\kappa\lambda} (\CD_{(\mu} \CD_{\rho} \CD_{\kappa)} \phi_1) \phi_2 a_3(\De) (\CD_{(\nu} \CD_{\sigma} \CD_{\lambda)}  \phi_3) \phi_4 \Big]
\, .
\end{aligned}
\end{equation}
The reason for this form of the functional is that each of the vertices carries at most three uncontracted derivatives. All other uncontracted derivatives that come from the propagator must be contracted in some way to form a \De{} acting either on a single $\phi$ which is eliminated using the equation of motion, or is absorbed in the one of the functions $a_i$. To arrive at the form \eqref{eq:amp_genform}, one has to use the commutator rules derived in appendix \ref{sec:commutators} together with partial integrations. With some effort, one arrives at the result
\begin{equation}
\begin{aligned}
 a_0(\De) &= 0 \, , \\
 a_1(\De) &= -(1+\De) \, , \\
 a_2(\De) &= \frac{2(2d+(d-1)\De)}{2+(d-1)\De}\, , \\
 a_3(\De) &= 0 \, .
\end{aligned}
\end{equation}
The calculation of the functions $a_i$ is the main result of this work. We've checked that the result doesn't depend on the gauge fixing parameters $\alpha, \{\beta_i\}$.

To make contact to standard quantum field theory notions of scattering in momentum space, it is useful to introduce curved-spacetime generalisations of the Mandelstam variables,
\begin{equation}\begin{aligned}
			\mans \,	(\phi_1 \phi_2	\phi_3	\phi_4)
	&\equiv
			\phi_1	\phi_2	\,	(\De \phi_3	\phi_4)
	\,\text{,}\\
			\mant	\, (\phi_1 \phi_2	\phi_3	\phi_4)
	&\equiv
			\phi_2	\phi_4	\,	(\De \phi_1	\phi_3)
	\,\text{.}
\end{aligned}\end{equation}
It should be noted that these operators do not commute in curved spacetime. Our convention is that all $\mant$-operators are sorted to the right.
All other contractions of derivatives can be eliminated by partial integration, the equation of motion and by using the commutator rules derived in \autoref{sec:commutators}. 

With these operators, the amplitude operator $\mathfrak A$, which is the operator equivalent of an amplitude in momentum space, reads
\begin{equation}\label{eq:amplitude}\begin{aligned}
			\mathfrak{A}
	&=
			4\alphag \left(\mans + \frac{2}{d-1}\right)^{-1} \times\\
	&\qquad
			\left(a_{01}	\mant
			+	a_{11}	\mans\mant
			+	a_{21}	\mans^2\mant
			+	a_{02}	\mant^2
			+	a_{12}	\mans\mant^2\right)
	\,\text{,}
\end{aligned}\end{equation}
where the coefficients $a_{ij}$ are given by
\begin{equation}
\begin{aligned}
			a_{01}	&=	\frac{2d+1}{1-d}
	\,\text{,}\qquad&
			a_{11}	&=	\frac{1}{1-d}	-	\frac{3}{2}
	\,\text{,}\\
			a_{21}	&=	-\frac{1}{2}
	\,\text{,}\qquad&
			a_{02}	&=	\frac{d}{1-d}
	\,\text{,}\\
			a_{12}	&=	-\frac{1}{2}
	\,\text{.}&&
\end{aligned}
\end{equation}

\subsection{Discussion}
The amplitude computed in \autoref{sec:result} possesses several striking features. First, the amplitude is gauge-invariant, as it must be since it is directly related to an observable, the cross section.

Secondly, we note that the flat limit is to be taken with great care. Using the equation of motion \eqref{eq:EEsol}, we find that the Mandelstam operators are related to the usual Mandelstam variables by $\mans \propto s/\CC$, $\mant \propto t/\CC$. The amplitude $A$, in the limit $\CC\to0$, has the leading order behaviour
\begin{equation}\label{eq:flatamplitude}\begin{aligned}
			A
			&\propto \alphag \left[ \frac{(s+t)t}{\CC^2} + \mathcal{O}(\CC^{-1}) \right]
	\\&\propto
			G_N \left[ {(s+t)t} \, \CC^{\frac{d-6}{2}}	 + \mathcal{O}(\CC^{\frac{d-4}{2}}) \right]
	\,.
\end{aligned}\end{equation}
Sub-leading contributions of $\mathcal{O}(\CC^{\frac{d-4}{2}})$ depend on the ordering of the operators, and are therefore non-universal.

The momentum scaling, and correspondingly the scaling with $\CC$, in the flat limit can also be derived by power counting of derivatives in the action. We observe that the propagator scales like $\propto E^{-2}$, where $E$ denotes the center-of-mass energy. The vertices each contribute a factor of $E^3$, yielding an amplitude $\propto E^4$. Since the amplitude is dimensionless, this has to be compensated by the only dimensionful quantity that we have at our disposal, namely $\CC$.

Specialising to $d=4$, we find that the amplitude \eqref{eq:flatamplitude} resembles the amplitude obtained in \GR{}, $A^{\text{\GR}} \propto \frac{(s+t)t}{s}$, except that the factor $1/s$ is replaced by $1/\CC$. The similarity is affirmed by performing a partial-wave analysis of \eqref{eq:flatamplitude}. The resulting spectrum contains contributions with spin 0 and spin 2, in agreement with \GR{}.

Since the amplitude diverges as $\CC^{-1}$, the flat space limit in $d=4$ is formally ill-defined.
Inserting the observed values of $\CC$ and $G_N$, we observe that the amplitude becomes of order unity around the meV scale. This implies that quantum gravitational effects would become relevant at this scale. Taking the scaling with energy at face value, this would rule out the theory experimentally.

Nevertheless, there are at least two reasons why this result might still be modified. First, using dimensional analysis, it is straightforward to deduce that loop diagrams of any order contribute to the flat limit. It is therefore conceivable that a resummation of loop diagrams will ameliorate the divergent behaviour.
Second, the amplitude can be modified by introducing non-minimal interactions \cite{Draper:2020bop, Draper:2020knh}. While this may not cancel the divergence in general, this could be achieved by a fine-tuning of the couplings, or be imposed automatically by a fixed point of the theory's \RG{}.

\section{Summary and outlook}\label{sec:summary}
In this paper, we have given a modern perspective on affine gravity. This theory is classically equivalent to \GR{}, but  its dynamical degrees of freedom are based on the connection rather than the metric. In affine gravity, the cosmological constant $\CC$ has the role of an integration constant, providing a possible explanation for its small observed value.

The main result of this work comprises the tree-level amplitude functional \eqref{eq:amp_genform} of a two-scalar-to-two-scalar process in a curved background, giving insight into the quantum degrees of freedom of the theory. Taking the flat-space limit, we find that in four dimensions, the amplitude scales like $\CC^{-1}$.
While this is worrying at first sight, we propose that the introduction of non-minimal interactions and loop corrections may ameliorate this behaviour. The computation of these modifications is left for future work. The calculation also shows that in the flat space limit, the off-shell gauge-invariant degrees of freedom of the theory have a spin two and a spin zero contribution, in agreement with metric gravity. At finite curvature, this structure appears to be more intricate.

A systematic search of non-minimal interactions can be done in terms of form factors, which can be successfully used to parameterise metric actions \cite{Knorr:2019atm, Draper:2020bop, Draper:2020knh}. In affine gravity, the occurrence of the inverse symmetric Ricci tensor $\invric^{\mu\nu}$ prohibits an expansion in powers of the curvature tensor. A first step in investigating interactions can be done in $d=3$ dimensions, where the Riemann tensor is fully determined by $\ric_{\mu\nu}$ only.

Due to the treatment of the affine connection as a standard gauge field, the affine formalism lends itself to making a relation to other approaches. First, affine gravity is straightforwardly generalised to include torsion. Second, treating the affine connection and gauge fields on equal footing, one could speculate about Grand Unification scenarios of gravity. Finally, affine gravity may allow for a lattice formulation, opening up new computational toolkits to investigate the quantum properties of this theory.

\section*{Acknowledgements}

We would like to thank Holger Gies, Daniel Litim, and Lee Smolin for interesting discussions, and Martin Reuter for helpful comments on the manuscript. B.\ K.\ acknowledges support by Perimeter Institute for Theoretical Physics. Research at Perimeter Institute is supported in part by the Government of Canada through the Department of Innovation, Science and Industry Canada and by the Province of Ontario through the Ministry of Colleges and Universities.

\appendix

\section{Commutators}\label{sec:commutators}

To sort the scattering amplitude into a canonical form, we need to know how we can commute functions of the operator \De{} with covariant derivatives. In alignment with the assumptions of the main text, we will consider a covariantly constant Ricci tensor,
\begin{equation}
 \CD_\mu \ric_{\alpha\beta} = 0 \, ,
\end{equation}
together with a maximally symmetric Riemann tensor of the form
\begin{equation}\label{eq:symriem}
 \riem{\mu\nu\rho}{\sigma} = \frac{1}{d-1} \left( \ric_{\mu\rho} \deltatensor{\nu}{\sigma} - \ric_{\nu\rho} \deltatensor{\mu}{\sigma} \right) \, .
\end{equation}
As a direct consequence, we find that
\begin{equation}
 \invric^{\mu\rho} \riem{\mu\nu\rho}{\sigma} = \deltatensor{\nu}{\sigma} \, .
\end{equation}
This equality indeed follows directly from a covariantly constant Ricci tensor without invoking \eqref{eq:symriem}, since
\begin{equation}
 0 = \invric^{\mu\kappa} \invric^{\lambda\tau} [\CD_\mu, \CD_\nu] \ric_{\kappa\lambda} = \deltatensor{\nu}{\tau} - \invric^{\kappa\lambda} \riem{\kappa\nu\lambda}{\tau} \, .
\end{equation}
The basic commutators are
\begin{equation}
 [ \CD_\mu, \CD_\nu ] X_\rho = \riem{\mu\nu\rho}{\alpha} X_\alpha = \frac{1}{d-1} \left( \ric_{\mu\rho} X_\nu - \ric_{\nu\rho} X_\mu \right) \, ,
\end{equation}
and
\begin{equation}
\begin{aligned}
 {} [ \CD_\mu, \CD_\nu] Y^\rho &= -\riem{\mu\nu\alpha}{\rho} Y^\alpha \\
 &= -\frac{1}{d-1} \left( \ric_{\mu\alpha} \deltatensor{\nu}{\rho} Y^\alpha - \ric_{\nu\alpha} \deltatensor{\mu}{\rho} Y^\alpha \right) \, .
\end{aligned}
\end{equation}
By the assumption that the Ricci tensor is covariantly constant, we have
\begin{equation}
\begin{aligned}
 {} [ \CD_\mu, \CD_\nu] Y^\rho &= [ \CD_\mu, \CD_\nu] \invric^{\rho\alpha} \ric_{\alpha\beta} Y^\beta \\
 &= \invric^{\rho\alpha} [ \CD_\mu, \CD_\nu]  (\ric \,Y)_\alpha \, ,
\end{aligned}
\end{equation}
thus it is enough to derive formulas for tensors with only lower indices.

We want to derive a formula of the form
\begin{equation}
 f(\De) \CD_\alpha \mathbf X = \CD_\alpha f(\De) \mathbf X + \ldots \, ,
\end{equation}
where $f$ is some arbitrary function and $\mathbf{X}$ is a tensor of rank $(0,n)$,
\begin{equation}
 \mathbf X = X_{\mu_1 \cdots \mu_n} \, .
\end{equation}
To derive this equation, we employ a standard trick and write the function $f$ as an inverse Laplace transform, so that we can use the Baker-Campbell-Hausdorff formula,
\begin{equation}\label{eq:fcomm}
\begin{aligned}
 f(\De) \CD_\alpha \mathbf X &= \int_0^\infty \text{d}s \, \tilde f(s) \, e^{-s\De} \CD_\alpha \mathbf X \\
 &= \int_0^\infty \text{d}s \, \tilde f(s) \, \sum_{\ell \geq 0} \frac{(-s)^\ell}{\ell!} [\De, \CD_\alpha]_\ell \, e^{-s\De} \mathbf X \, .
\end{aligned}
\end{equation}
Here, we used the multi-commutator, which is defined recursively by
\begin{equation}
 [A,B]_n = [A, [A,B]_{n-1}] \, , \qquad [A,B]_0 = B \, .
\end{equation}
To make progress, let us first calculate the standard commutator. Using the standard rules for commutators and our assumptions of covariant constancy and maximal symmetry, we find
\begin{widetext}
\begin{equation}
\begin{aligned}
 {}[\De, \CD_\alpha] X_{\mu_1 \cdots \mu_n} &= - \CD_\alpha X_{\mu_1 \cdots \mu_n} - \frac{2}{d-1} \sum_{k=1}^n \left[ \CD_{\mu_k} X_{\mu_1 \cdots \mu_{k-1} \alpha \mu_{k+1} \cdots \mu_n} - \invric^{\kappa\lambda} \ric_{\alpha\mu_k} \CD_\kappa X_{\mu_1 \cdots \mu_{k-1} \lambda \mu_{k+1} \cdots \mu_n} \right] \\
 &\equiv \Ctensor{\alpha\mu_1 \cdots \mu_n}{\beta \nu_1\cdots\nu_n} \CD_\beta X_{\nu_1 \cdots \nu_n} \, .
\end{aligned}
\end{equation}
\end{widetext}
Here, we made the observation that the commutator is in fact a multiplication with a covariantly constant tensor $\mathbf{\Ctensor{}{}}$, so that structurally we find,
\begin{equation}
 [\De, \CD]_k \mathbf X = \mathbf{\Ctensor{}{}}^k \CD \mathbf X \, .
\end{equation}
Plugging this into the original equation \eqref{eq:fcomm}, we find
\begin{equation}
 f(\De) D \mathbf X = \int_0^\infty \text{d}s \, \tilde f(s) \, \text{Texp}\left[ -s \, \mathbf{\Ctensor{}{}} \right] \, \CD \, e^{-s\De} \mathbf X \, .
\end{equation}
Here, Texp is the tensor exponential, defined in terms of its power series.

While the calculation of this exponential for an arbitrary rank tensor seems difficult, we can easily calculate it for tensors of low rank, and in practice, this is all that is needed. We will illustrate this for scalars and vectors. For a scalar,
\begin{equation}
 [\De, \CD_\alpha] X = - \CD_\alpha X \, ,
\end{equation}
so that
\begin{equation}
 \Ctensor{\alpha}{\beta} = -\deltatensor{\alpha}{\beta} \, .
\end{equation}
Correspondingly, we have
\begin{equation}
 f(\De) \CD_\alpha X = \CD_\alpha f(\De-1) X \, .
\end{equation}
This emphasises again a structural difference to standard metric-based calculations - since \De{} is dimensionless, such commutator formulas involve shifts by constants instead of shifts by curvatures. In complete analogy, for a vector, we have
\begin{equation}
 \Ctensor{\mu\nu}{\alpha\beta} = -\deltatensor{\mu}{\alpha} \deltatensor{\nu}{\beta} - \frac{2}{d-1} \deltatensor{\mu}{\beta} \deltatensor{\nu}{\alpha} + \frac{2}{d-1} \ric_{\mu\nu} \invric^{\alpha\beta} \, ,
\end{equation}
which gives
\begin{equation}
\begin{aligned}
 f(\De) &\CD_\mu X_\nu = \CD_{(\mu} f\left(\De - \frac{d+1}{d-1}\right) X_{\nu)} \\
 & + \frac{1}{d} \ric_{\mu\nu} \invric^{\alpha\beta} \CD_\alpha \left[ f\left( \De + 1 \right) - f\left(\De - \frac{d+1}{d-1} \right) \right] X_\beta \\
 & + \CD_{[\mu} f\left(\De - \frac{d-3}{d-1}\right) X_{\nu]} \, .
\end{aligned}
\end{equation}
We shall not give the formula for a rank two tensor, since it is rather lengthy. The formula can be found in the Mathematica notebook attached to this work.

\bibliography{general_bib}

\end{document}